# Translational and rotational temperatures of a 2D vibrated granular gas in microgravity.

Y.Grasselli[(1)(2)], G.Bossis[(1)], R.Morini[(1)]

[(1)] LPMC UMR6622 - University of Nice – Parc Valrose – 06108 Nice Cedex 2 (France)

[(2)] SKEMA Bachelor – 60 rue Dostoievski – BP085 – 06902 Sophia Antipolis (France)

*Abstract*

We present an experimental study performed on a vibrated granular gas enclosed into a 2D rectangular cell. Experiments are realized in microgravity. High speed video recording and optical tracking allow to obtain the full kinematics (translation and rotation) of the particles. The inelastic parameters are retrieved from the experimental trajectories as well as the translational and rotational velocity distributions. We report that the experimental ratio of translational versus rotational temperature decreases to the density of the medium but increases with the driving velocity of the cell. These experimental results are compared with existing theories and we point out the differences observed. We also present a model which fairly predicts the equilibrium experimental temperatures along the direction of vibration.

**Introduction**

Granular gases display a much more complex behavior than molecular gases like anisotropy of temperature along different directions [1, 2], coexistence of different temperatures depending on the size of the particles [3, 4], non Gaussian distribution of particle velocity [5, 6] or cluster formation at high enough density [7, 8]. One of the major aspects of these flows is the dissipative nature of granular material and the dynamics of such systems finds its origin in the binary collisions between particles. A granular material requires continuous input of energy for a sustained flow. The amount of the heat flux injected at the boundaries strongly influences the flow of granular materials especially in the case of vibrated beds. Most of the studies of vibrated granular media focus on the prediction of the granular temperatures and the velocity distribution functions along the direction of vibration and perpendicular to it [1, 6, 9] but neglect the coupling with the rotational degrees of freedom.

Previous works have been done on the collisional properties of particles. In its basic definition, a collision is assumed to be instantaneous and the inelasticity is described by a normal restitution coefficient. However, the particles of a granular medium are not perfectly smooth and surface roughness play also a role during collision providing a transfer of angular momentum between particles leading to the rotations of the grains. Thus, a full description of a collision requires the knowledge of the normal and tangential restitution coefficient. Consideration of the rotation of the particles requires the determination of the tangential restitution coefficient which is not easy to realize experimentally since it is needed to track the rotation of the particles with a high speed camera and with some marks printed on the surface of the particles [10]. On the other hand numerical simulations do not suffer from this experimental constraint and several simulations deal with the rotational component and compare their results mainly with kinetic theories [11-13].

Our aim, here, is to provide experimental data both for the normal and tangential restitution coefficients and for the different quantities related to the rotational and translational degrees of freedom such as the distribution functions and the rotational and translational temperatures. All of these being obtained directly from the kinematics of granular particles submitted to a vertical vibration. We shall particularly focus on the ratio between rotational and translational temperatures. In order to remove the gravity bias, we have conducted the experiments in a low gravity environment. Several other groups have already presented experimental results on granular flow under such conditions [14-16] but to our knowledge this is the first experiment giving access to rotational and translational velocities and so, the corresponding temperatures.

In the next section we first describe our experimental set-up, the type of model particles used and the conditions of the experiment. We will explain how the properties of particles as well as their kinematics are obtained from a direct optical tracking and analysis of their trajectories. The experimental results on temperatures and velocity distributions will also be given. In the last section our experimental results will then be compared to existing theories considering the coupling between translational and rotational motion and we shall discuss the way that the translational granular temperature can be predicted from these models.

**Experiments**

In order to investigate the dynamical behavior of a model granular medium, we have designed a 2D-cell of rectangular shape, with a height $d_v = 6.8 cm$ and a width $d_h = 6 cm$, wherein brass disks having a diameter $\sigma = 6mm$ and mass $m = 4.6\ 10^{-4} kg$ are enclosed between two glass plates. The initial area fraction $\varphi$ of the medium is given by the number of disks, $N$, into the cell, here 12 or 24 disks, corresponding respectively to area fractions of 8.3% and 16.6%. The cell is mounted on a vibrating device to submit the medium to a periodic (sine oscillations) external vibration with different frequencies, $\nu$, and amplitudes $A$ (fig. 1). The vibration is applied along the $y-$direction (which is the direction of normal gravity). To cancel gravitational effects, all experiments have been performed in microgravity: the experimental apparatus is boarded in the airplane A-300 Zero G from Novespace. The airplane undergoes successive parabolic flights allowing around 22s of microgravity per parabola. Note that the vibration is already present when the plane enters microgravity in order to leave enough time to the system to reach the thermal equilibrium. During each microgravity sequence, high speed video recordings are realized on the vibrated granular medium to obtain the trajectories of the particles. To reduce friction effects between the disks and the glass plates of the cell, each disk is dressed, on each of its side, by three small steel beads: this configuration also reduces the tilting of the disks in the presence of the external vibration between the cell's plates. The aim of this study is to obtain, by direct optical tracking, the kinematics of the granular particles and thus to access all parameters involved in the dynamics of the medium. To achieve such goal, each disk is pierced with two small holes, symmetric about the center of the disk and video observations are realized by light transmission (fig. 2). Images have a resolution of 720 x 720 pixels and the frame rate is 900 FPS: we thus record about 22000 images during each parabola. It grants us with images having a high contrast and quality and allows individual tracking by direct image analysis [17]. To determine the position of the disk, we track the trajectories of the two holes of each disk as a function of time. The barycenter then gives the $x-$ and $y-$position of the disk, from which the linear components of the velocity $v_x(t)$ and $v_y(t)$ can be computed. Moreover, the determination of the time dependence of the angle $\theta(t)$, computed through the angular position of the holes from the horizontal direction, gives access to the angular velocity $\dot{\theta}(t)$. Since the two holes are well identified during tracking, the orientation angles of the disks are fully determined from 0 to 360 degrees. One can observe in figures 3 and 4 a typical experimental record of $\theta(t)$ and of the trajectory of a particle within the cell. On figure 4, it is interesting to note that a sharp change in the direction of rotation or a significant variation of the slope, both result from a collision with another particle. On the contrary, when the particle experiences no collision (e.g. time larger than 5s in figure 4), the angular velocity remains quite constant, indicating the absence of friction with the lateral walls. As mentioned above, precautions to reduce friction effects arising between the disks and the lateral walls of the cell were taken but we have sometimes observed the presence of this undesired effect along some parts of the disks' trajectories. This is due either to the motion of the disk but also from small gravity fluctuations in the direction normal to the lateral walls. Nevertheless, during an experiment, it is easy to identify from the trajectories of the disk, the time intervals where friction becomes non negligible and to remove them from the experimental processing used to determine the inelastic parameters. This issue being rarely encountered, in the following we will neglect friction with the lateral walls of the cell in the experimental determination of the temperatures and densities.

The analysis of the trajectory of each disk allows a systematic investigation of inelastic properties of the particles: normal, $r$, and tangential, $\beta_0$, restitution coefficient. By tracking the change in direction of motion of each disk when a nearest neighbor is present, we are able to precisely determine the binary collisions from the trajectories. If both conditions are satisfied, we then know the time $t_c$ at which a collision arises. For the two disks involved in a collision, we consider the previous and following positions from $t_c$ (fig. 5). The positions considered must insure that the trajectories before and after collision are linear: the determination of the exact position of the disks at collision and the direction of the normal direction $n$ is then possible. From the experimental linear and rotational velocities measured before and after collision, we can obtain the inelastic parameters. We compute the relative velocities before impact, $\vec{V_R}$, and after impact, $\vec{V'_R}$. The general expression of the relative velocity can be written as $\vec{V_R} = \vec{v_1} - \vec{v_2} - a\left(\dot{\vec{\theta}}_1 + \dot{\vec{\theta}}_2\right) \times \vec{n}$ where the subscripts 1 and 2 stand for the two colliding particles at a given time (fig. 6). Normal and tangential coefficients are then obtained from: $r = -\left|\vec{n} \cdot \overrightarrow{V'_R}\right|/\left|\vec{n} \cdot \overrightarrow{V_R}\right|$ and $\beta_0 = -\left|\vec{n} \times \overrightarrow{V'_R}\right|/\left|\vec{n} \times \overrightarrow{V_R}\right|$. On the other hand, if we introduce the angle $\gamma$ between $\vec{n}$ and $\overrightarrow{V_R}$, we have for disks the relation [12]: $1 + \beta_0 = -3(1 + r)\mu \cot(\gamma)$. The initial slope of $\beta_0$ versus $\cot(\gamma)$ allows the computation of, $\mu$, the friction coefficient.

We obtained experimentally an average value of $r = 0.64 \pm 0.03$. Despite it is sometimes noticed in such situation [18-19], we did not observe in our experiments any clear dependence of $r$ on the relative impact velocity. The experimental determination of the restitution coefficient, $\alpha$, between a particle and the walls of the cell report a value $\alpha = 0.71 \pm 0.04$. We were also able to determine the behavior of the experimental tangential restitution coefficient as a function of $cot(\gamma)$. The results are presented in figure 7. From the initial slope one can compute an average value for the friction coefficient during a binary collision: $\mu = 0.14 \pm 0.01$. Due to the specific shape of the cell and our experimental conditions, most of the binary collisions are taking place for large values of $cot(\gamma)$ so the average value of the tangential restitution can be taken as $\beta_0 = 0.7 \pm 0.05$.

The density and local velocity profiles of particles within the cell can be determined again from the positions of particles. A typical result of local area fraction along the direction of vibration as well as the corresponding velocity profile are reported in figure 8. It is clearly observed that the regions close the top and bottom walls of the cell show a low concentration of particles while in the center, the area fraction of particles is almost two times the initial one. This result is a direct consequence of inelastic collisions which tend to form clusters of particles [20]. As proposed by R. Soto [21], the cell may be divided into two different parts: one at the center of the cell that we will referred as "cold" and two ones close to the top and bottom walls, named as "hot" where energy is injected into the medium. To define the size, $H_H$, of the "hot" regions, we have used, for each experiment, the position of the intersection of the density and average velocity profiles. Considering all experiments, we have noticed first, that the values found for $H_H$, were not really sensitive to the amplitude of vibration as one could expect, second that an average value of $H_H = 9mm \approx 1.5\,\sigma$ was acceptable for all the experiments performed.

The temperature of the granular medium can be computed from the velocities of particles, including the translational, $T_{tr}$, and the rotational temperatures $T_{rot}$. In a steady state, the equilibrium temperature of the medium is given by the balance between the energy flux injected into the medium through the collisions of particles with the top and bottom walls of the cell (i.e. in the "hot" regions) and the energy flux dissipated in the bulk (i.e. the "cold" region) due to inelastic collision between particles. xEnergy injection is then performed in the top and bottom areas of the cell while the main energy dissipation occurs in the central area. Note that all experimental temperatures obtained in the following have been measured in the "cold" zone. Moreover, from the density profiles, it is possible to determine an average number, $N_H$, of particles present in the "hot" regions of the cell at any time. Finally, the velocity distributions are also obtained from the kinematics of particles. Both typical distributions for translation and rotation velocities are presented respectively on figure 9 and 10. A clear Maxwell-Boltzmann behavior is observed. The dashed line on the figures represent the plots of the theoretical expression of the distribution in which the experimental value of the squared velocities has been introduced.

Due to the rectangular shape of the experimental cell used, and to the relatively low area fraction, the main contribution to the temperature was expected to be found along the direction of the external vibration (the $y$ −direction). In the following chart we present a summary of the temperature ratios $T_y/T_x$ and $T_{tr}/T_{rot}$ with $T_{tr} = (T_x + T_y)/2$, in terms of the maximum cell's velocity $A\omega$ ($\omega = 2\pi\nu$) for the two area fractions used.

| | $A\omega(m/s)$ | 0.212 | 0.22 | 0.283 | 0.380 | 0.442 |
|---|---|---|---|---|---|---|
| | $A(m)$ | 5.6 10$^{-4}$ | 2.3 10$^{-3}$ | 3 10$^{-3}$ | 2 10$^{-3}$ | 4.7 10$^{-3}$ |
| $\varphi = 16.6\%$ | $T_y/T_x$ | 1.98 | 3.184 | 3.51 | 2.96 | 4.00 |
| | $T_{tr}/T_{rot}$ | 6.40 | 4.49 | 6.50 | 9.01 | 9.60 |
| | $A\omega(m/s)$ | 0.22 | 0.30 | 0.347 | 0.417 | 0.556 |
| | $A(m)$ | 3.5 10$^{-3}$ | 4.7 10$^{-3}$ | 3.7 10$^{-3}$ | 4.4 10$^{-3}$ | 6 10$^{-3}$ |
| $\varphi = 8.3\%$ | $T_y/T_x$ | 5.47 | 5.88 | 5.92 | 5.27 | 7.11 |
| | $T_{tr}/T_{rot}$ | 11.73 | 13.70 | 15.25 | 14.24 | 24.33 |

*Table 1: ratios of temperatures for two different area fractions: $\varphi = 16.6\%$ and $\varphi = 8.3\%$ for different driving velocities $A\omega$ and amplitude vibrations A.*

We shall analyze these experimental results by focusing first on the ratio $T_y/T_x$ which is clearly dependent on the area fraction of the medium and is larger for the smallest area fraction. The temperatures found along the direction of vibration are always larger than the ones in the transverse direction which is not a surprising result since the main part of energy injection is performed along the $y$ −direction and the relatively low area fraction does not allow to redistribute this energy on the perpendicular direction. At low area fraction, the particles can move easily and the $y$ −direction drives the general motion. On the other hand, we also observe a net increase of the ratio $T_y/T_x$ with the driving velocity of the cell, but less pronounced for the lower area fraction. However, the driving velocity is not the only parameter of the problem and the amplitude can also play a role. For example, the ratio $T_y/T_x = 1.98$ reported was achieved with the smallest amplitude ($A = 0.556mm$) and the largest frequency ($60Hz$). For these experimental conditions (large frequency and small amplitude), we observe that the particles mainly concentrate in the center of the cell and consequently the energy injection through the moving walls of the cell is weak. This might explain the low ratio obtained in this experimental run, compared to the one with almost the same value of $A\omega = 0.22 m/s$ but a much larger amplitude: $A = 2.3 mm$. For all the other results, the frequency is in between $10 Hz$ and $30 Hz$ and corresponding amplitudes of vibration are large enough to continuously shake the medium.

The second result is related to the ratio $T_{tr}/T_{rot}$ which clearly increases with $A\omega$ and which also depends strongly on the volume fraction of the medium. The translational temperature is about one order of magnitude larger than the rotational temperature. Again, the fact that most of the collision are quite head-on ones in this geometry, as reflected by the high value of $T_y/T_x$ may explain why the transfer from translational to rotational energy is rather weak, especially at the lowest area fraction.

**Comparison with existing theories**

In a mean field theory, the rate of change of the temperature of a granular medium is determined through two coupled equations [13]:

$$\begin{cases} \frac{dT_{tr}}{dt} = H_{dr} + G\left[-AT_{tr}^{3/2} + BT_{tr}^{1/2}T_{rot}\right] \\ \\ \frac{dT_{rot}}{dt} = 2G\left[B'T_{tr}^{3/2} - CT_{tr}^{1/2}T_{rot}\right] \end{cases} \qquad (1)$$

Where $T_{tr}$ and $T_{rot}$ represent respectively the translational and rotational temperatures and $G = \frac{8}{a\sqrt{\pi m}}\varphi g_2(\varphi)$ is related to the collision rate between particles; $g_2(\varphi)$ being the pair correlation function at contact. In 2D, $g_2(\varphi) = (1 - 7\varphi/16)/(1 - \varphi)^2$. The constants $A, B, B'$ and $C$ depend only on the inelastic properties of the particles (more details are given in [13]). $H_{dr}$ is for the energy input into the medium and, in this analysis, the energy is supposed to be injected homogeneously into the medium. Note that these constants are positive so that the minus signs express the dissipative behavior of the medium.

Several inelastic modelizations were proposed by Herbst *et al.* ranked from "model A" to "model E" [13]. We briefly report the different models: "Model A" considers a constant tangential restitution coefficient. "Model B" considers a mean tangential restitution coefficient calculated from a simplified probability distribution of the impact contact angle: $P(\gamma) = -\cos(\gamma)$ whereas in "Model C" the distribution $P(\gamma)$ is computed analytically and is in good agreement with the simulation results. Finally, "Model D" and "model E" are obtained respectively with a tangential restitution depending on $\gamma_{12}$ (the contact angle obtained neglecting the rotational velocities) or on the real contact angle, $\gamma$.

From the second equation of (1), the energy ratio $T_{tr}/T_{rot}$ can be obtained considering the medium at thermal equilibrium, $dT_{rot}/dt = 0$, allowing to get the relation $T_{tr}/T_{rot} = C/B'$. Depending on the model used, the expressions of the constant $C$ and $B'$ are given and only related to the inelastic properties of particles and to their inertia but neither to the area fraction nor to the driving energy flux $H_{dr}$. Introducing the values of the normal and tangential restitution and friction coefficient from our experiments, gives the following results.

| Model | A | B | C | D | E |
|---|---|---|---|---|---|
| $T_{tr}/T_{rot}$ | 1.53 | 3.23 | 3.87 | 3.73 | 5.2 |

*Table 2: Ratio of translational to rotational temperature for the different models proposed in ref. [13]*

Although these results are lower than the experimental ratio found, the model which better fits is, as expected, the more detailed one (i.e. "Model E"). Note that the predictions are identical for the two area fractions since the coefficients C and B' are only dependent of the restitution coefficients, whereas we have a strong difference with the area fraction from experimental results. Also the model does not predict a dependence with $A\omega$ which is not consistent with our experimental observations. Actually, these models do not deal with an anisotropic temperature since in the simulations, the energy is injected in an isotropic way. This is likely the main reason for the non-ability of these models to represent our experimental results

Next, we focus on the equilibrium temperature of the medium. When submitted to the external vibration, the medium can be modeled as a dissipative medium to which a given amount of energy is injected through the vibration per unit time. The equilibrium temperature is obtained by solving the equilibrium equation $H_{dr} + Q_d = 0$, where $H_{dr}$ is the energy flux injected in the medium by the collisions of particles with the walls of the cell and $Q_d$, the energy flux dissipated during the binary collisions between particles. $H_{dr}$ takes place in the regions close to the top and bottom walls, while $Q_d$ is determined in the bulk of the medium. The experimental results obtained with our cell's geometry clearly show that the main part of the energy of the particles is distributed along the direction of the external vibration ($y$ −direction). Based on experimental observations, we define the regions of energy injection by two layers of thickness $H_H$ close to the top and bottom moving walls and having the same width, $d_h$, of the cell. In these two regions, the density of particles is smaller than the average density of the medium; we call $N_H$ the average number of particles present at any time in this region. Thus, the bulk of the medium (i.e. the "cold" zone) reduces to dimensions $H_C = d_v - 2H_H$ where only $N_C = N - 2N_H$ particles are present at any time (subscripts $C$ and $H$ stand for "cold" and "hot"); the surface of this zone is then $S_C = H_C d_h$.

In the "cold" zone, the dissipated energy depends on the collision frequency $f_E(T)$ which in turns depends on the temperature $T$ of the medium, $T = m\langle v_x^2 + v_y^2\rangle/2$. If we neglect the loss of energy coming from tangential restitution coefficient, the energy dissipated per collision is given by:

$$\Delta E_{pp} = m\frac{(r^2-1)}{4}\langle[(\vec{v_1} - \vec{v_2}) \cdot \vec{n}]^2\rangle = \frac{(r^2-1)}{2}T \tag{2}$$

The frequency collision which is the inverse of the Enskog time is given in 2D by [22]:

$$f_E = \sqrt{2\pi}\frac{N_C}{S_C}\sigma g_2(\varphi)\langle v\rangle = \frac{2}{N_C}f_E^N \tag{3}$$

where $N_C/S_C$ represents the number density in the "cold" region and $f_E^N$ is the number of collisions between $N$ particles per unit time. Finally the dissipated energy flux will be (see also Appendix in [13]):

$$Q_d = f_E^N \Delta E_{pp} = \frac{N_C^2}{H_C d_h}\frac{1-r^2}{2}\sigma g_2(\varphi)\sqrt{\frac{\pi}{m}}T^{3/2} \tag{4}$$

Since the temperature is anisotropic we have to replace in (4) $T$ by $(T_x + T_y)/2$ or $(1 + 1/R_T)T_y/2$, where $R_T = T_y/T_x$ so that instead of (4) we get:

$$Q_d = \frac{N_C^2}{H_C d_h}\frac{1-r^2}{4}\sigma g_2(\varphi)\sqrt{\frac{\pi}{2m}}T_y^{3/2}(1 + \frac{1}{R_T})^{3/2} \tag{5}$$

We now have to express the flux of the injected energy during the collisions between the particles and the top and bottom walls of the cell. During one collision, the change in kinetic energy of one particle is: $\Delta E_{pw} = m(v'^2_y - v_y^2)/2$ with $v'^2_y$ and $v_y^2$, respectively, are the velocity of the particle after and before collision with the cell's wall. The cell is assumed to move with a velocity $V_{dr}$.

The relative velocity equation gives $v' - V_{dr} = \alpha(V_{dr} - v)$ where $\alpha$ is the normal restitution coefficient between the particle and the wall. The change in kinetic energy of one particle may be rewritten as:

$$\Delta E(v_y, V_{dr}) = \frac{m}{2}\left[(1+\alpha)^2 V_{dr}^2 - 2(1+\alpha)V_{dr}v_y - v_y^2(1-\alpha^2)\right] \tag{6}$$

The energy flux, $h_{dr}$, associated with particles going towards the wall, can be expressed as:

$$h_{dr} = \frac{N_H}{2H_H}v_y \Delta E(v_y, V_{dr}) \tag{7}$$

where we have assumed that $N_H/2$ particles are going towards the wall. The net energy flux for a given wall velocity is then obtained by averaging the flux of the incoming particles with the velocity distribution function, $f(v_y)$ associated with the "cold" region and integrating on the velocities directed towards the wall:

$$H_{dr}(V_{dr}) = \int_0^\infty h_{dr} f(v_y) dv_y \tag{8}$$

The velocity distribution is intended for particles which are going to collide with the top and bottom walls of the cells. We need to consider here the particles issued from the "cold" zone and having an average velocity directly related to the average temperature measured in this area of the cell. The dashed curve (cf. fig. 9) represents a Maxwell-Bolzmann distribution

$$f(v_y) = \sqrt{\frac{1}{2\pi \langle v_y^2 \rangle}} e^{-v_y^2/\langle v_y^2 \rangle} \qquad (9)$$

where the average value of the velocity is the one retrieved from experiments. The matching is in very good agreement and we will consider this type of behavior in the following.

The integral (8) over the velocities gives the following result:

$$H_{dr}(V_{dr}) = \frac{m}{4} \frac{N_H}{H_H} [(1+\alpha)^2 V_{dr}^2 I_1 - 2(1+\alpha) V_{dr} I_2 - (1-\alpha^2) I_3] \qquad (10)$$

where $I_1$, $I_2$, and $I_3$ are the integrals $\int_0^\infty v_y^i f(v_y) dv_y$ ($i = 1..3$) which are respectively given by:

$$I_1 = \sqrt{\frac{T_y}{2\pi m}} \qquad I_2 = \frac{T_y}{2m} \qquad I_3 = \left(\frac{T_y}{m}\right)^{\frac{3}{2}} \sqrt{\frac{2}{\pi}} \qquad (11)$$

In this derivation all the particles are supposed to go from the bulk towards the wall, so we have neglected the double collisions (with the particle going away from the wall and hit a second time by the wall) that we did not observe in our video records. It remains to average on the wall velocity. Then the linear term in $V_{dr}$ cancels and the term in $V_{dr}^2$ averages to $(A\omega)^2/2$ which gives the following flux for the injected energy (after multiplying by 2 for the two walls):

$$H_{dr} = m \frac{N_H}{2H_H} \left[ (1+\alpha)^2 (V_{dr})^2 \sqrt{\frac{T_y}{2\pi m}} - (1-\alpha^2) \left(\frac{T_y}{m}\right)^{\frac{3}{2}} \sqrt{\frac{2}{\pi}} \right] \qquad (12)$$

If $\alpha = 1$ (perfectly elastic walls), we recover the expression given by Soto [21] for a sinusoidal vibration taking for their function $q(T/m(A\omega)^2)$ the constant value $q = \sqrt{2/\pi} = 0.8$ which is actually a very good approximation in the range of our experimental values of $T/m(A\omega)^2$.

The equilibrium between injection (Eq. (12)) and dissipation (Eq. (5)) gives:

$$T_y = \frac{\frac{N_H}{2H_H}(1+\alpha)^2}{\frac{N_C^2}{H_C d_h} \pi \sigma g_2(\varphi) \frac{1-r^2}{2} \left(1 + \frac{1}{R_T}\right)^{3/2} + 2\frac{N_H}{H_H}(1-\alpha^2)} m(A\omega)^2 \qquad (13)$$

The temperature is proportional to the square of the amplitude of the driving velocity. Since we know, for each experiment, the respective density of the "cold" and hot "domains" we can compare the theoretical predictions of Eq. (13) with the experimental values of $T_y$ calculated in the "cold" domain. In order to take into account the dissipation due to the tangential restitution coefficient, $\beta_0$, we use instead of $r$ in Eq. (13) an effective restitution coefficient $r_e$ proposed by S. McNamara and S. Luding [23]:

$$r_e = \sqrt{r^2 - \frac{q(1-\beta_0^2)}{1+2q-\beta_0}} \qquad (14)$$

Using $q = 0.5$ for a disk, $\beta_0 = 0.7$; we obtain $r_e = 0.462$ instead of $r = 0.64$. The comparison between the theoretical temperatures $T_y$ obtained from Eqs (13)-(14) with the experimental ones calculated in the "cold" region is presented in the figure 11. We observe a quite good agreement for the two volume fractions we have used. In order to have a useful prediction of the relation between the temperature and the driving velocity, the main point would be to be able to predict the density $N_H/H_H$ close to the wall instead of taking this value from the experimental profile as we have done in this work.

The anisotropy of the temperatures produced by a vibrating wall is scarcely studied in the literature. One can find a recent experimental study in a 3D-cylindrical configuration [24] where the anisotropy $R_T = T_y/T_x$ is reported versus the volume fraction of particles and is shown to increase strongly for volume fraction below 10% but remains smaller than our values. A theoretical analysis is presented in [25] based on two different Maxwellian distributions for the directions parallel and perpendicular to the vibration and a density along the vibration axis, $z$, proportional to $\exp(-\frac{mgz}{T})$. A balance between energy fluxes along and perpendicular to the direction of vibration gives the ratio $R_T$ and predicts that, for perfectly reflective side walls, this ratio would only depend on the restitution coefficient. This is clearly not the case in our experiments (cf. table 1) where the ratio $R_T$ is much larger at the lower density. It is not possible to directly transpose this theory to our experiments since our density profile is very different from a gravity driven one, but it may be possible to predict $R_T$ along the same lines as in [25] if we suppose a constant density in the "cold" zone.

**Conclusion**

We have conducted two-dimensional experiments with a vibrated granular gas in microgravity. From the video recording of the trajectories, we were able to obtain the translational and rotational trajectories of each particle. These trajectories were then used to deduce the kinetic parameters of the disks like the normal and tangential restitution coefficients, the friction coefficient, and all the information related to the distribution of velocities and density. In particular we have reported the translational temperatures along and perpendicular to the direction of vibration, and also the rotational temperatures. When compared to existing theories, it appears that there are important differences since even the full model predicts a too small ratio $T_{tr}/T_{rot} = 5.2$ instead of a value between 6 and 10 depending on the driving velocities. The difference is still higher for the lower volume fraction and neither the area fraction nor the driving velocities appear in the model which is based on a homogeneous gas of constant density. In a granular gas driven by the vibration of the cell there are two major differences with the assumptions of the model: first the density is not homogeneous and second the translational velocities are much higher in the direction of vibration than perpendicular to it. We have found that a balance of the energy fluxes along the direction, y, of vibration can represent fairly well the evolution of the temperature $T_y$ with the driving velocity and with the volume fraction. In this balance it is necessary to take into account the existence of two domains, one "hot" region with a low density and a "cold" region with a high density and also the contribution of the tangential velocities to the dissipation. At least the distinction between the dissipation due to the collisions between the particles which is proportional to the average temperature $T = (T_x + T_y)/2$ and the driving flux, which depends only on $T_y$, was introduced, but on the basis of the experimental ratio $T_y/T_x$. This ratio increases when the volume fraction decreases and it also depends on the driving velocity. A theoretical determination of $T_y/T_y$ which could reproduce these behaviors should involve the non-elastic collisions with the lateral walls, but is let for a future work.

**Acknowledgements**

We would like to thank the NOVESPACE and the CNES for giving us the possibility to board the A300-zero G in order to perform our experiments.

**Caption of the figures**

**Figure 1**: Sketch of the experimental configuration to track the motion of the disks using high speed video recording. The cell is mounted on a vibrating device allowing different amplitudes and frequencies. The disks are pierced with two holes and a light source, placed behind the cell, gives clear observations of the disks by light transmission.

**Figure 2**: Typical raw experimental picture recorded during the period of microgravity in the presence of the external vibration (along the *y*-direction). The two holes, used for the optical tracking of the particles, can be clearly identified. A side and top sketch of one disk is also shown. Three small steel beads are placed on both sides of each disk to reduce friction effect on the lateral walls of the cell and to prevent disk's tilting during vibration.

**Figure 3**: Experimental *(x,y)*-trajectory of one disk recorded in the presence of microgravity and external vibration. The inset shows how, from the knowledge of the positions of the two holes, the position of the disk and its orientation angle, $\theta$, can be determined.

**Figure 4:** Experimental recording of the angle of orientation, $\theta$, of one disk as a function of time in the presence of microgravity and external vibration. A sharp change in direction of rotation or slope indicates a collision with another particle. When no collision is encountered, the angular velocity remains almost constant (like on the right part of the curve).

**Figure 5**: Experimental trajectories recorded during a collision between disks. The circles represent the positions retrieved from optical tracking. For a better understanding, we have added on the experimental trajectories the direction of motion of the disks (arrows) before and after collision. We can precisely obtain the position of each disk at impact but also the direction of the normal direction *n* used in the determination of the inelastic properties.

**Figure 6**: Sketch of two colliding particles. $\vec{v_1}$ and $\vec{v_2}$, and, $\vec{\dot{\theta}_1}$ and $\vec{\dot{\theta}_2}$ represent, respectively the linear and rotational velocities of the particles before and after impact. $\vec{V_R}$ is the relative velocity and $n$ the normal direction at collision. The impact angle $\gamma$ is defined from $n$ to $\vec{V_R}$.

**Figure 7**: Experimental tangential restitution $\beta_0$ as a function of $cot(\gamma)$. This coefficient is calculated from binary collisions with, in addition, the knowledge of the angular velocities of the two disks before and after collision. The plain curve is a linear regression used to compute the friction coefficient arising between two particles at contact.

**Figure 8**: Experimental average density (circles) and velocity (triangles) profiles obtained along the direction of vibration in microgravity for an area fraction $\varphi = 16.6\%$. The particles are mainly located at the center of the cell and the density drops near the moving walls of the cells located on top and bottom. These profiles are used to determine the width of the area where energy is injected into the medium (named as "hot" zone, see text).

**Figure 9**: Velocity distribution of the component along the direction of vibration (experiment: plain curve). The dashed line corresponds to the mathematical plotting of a Maxwell distribution which includes the average velocity determined experimentally.

**Figure 10**: Typical angular velocity distribution of the particles (experiment: plain curve). The dashed line corresponds to the mathematical plotting of a Maxwell distribution which includes the average angular velocity determined experimentally

**Figure 11**: Comparison of the equilibrium temperature computed from equation 13 as a function of the driving velocity of the cell ($V_{dr}$). Theory and experiments are in good agreement for the two volume fractions used in experiments. The results show a clear dependence on the volume fraction of particles.

**Figure 1**

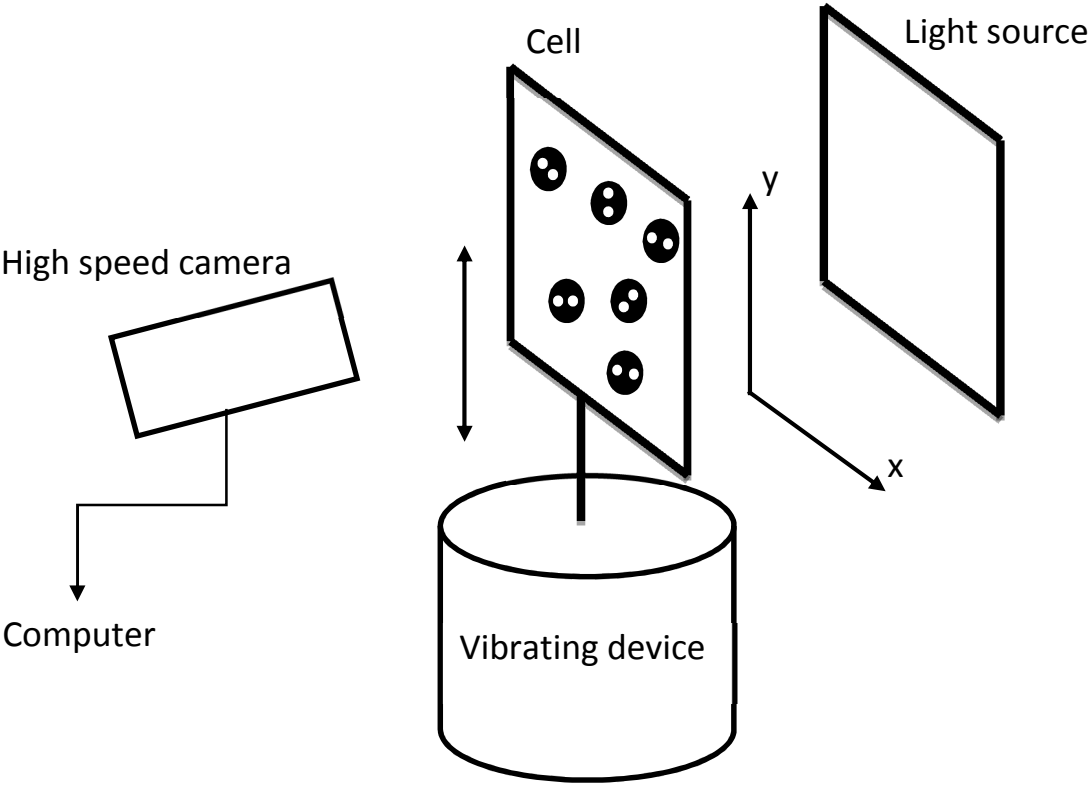

**Figure 2**

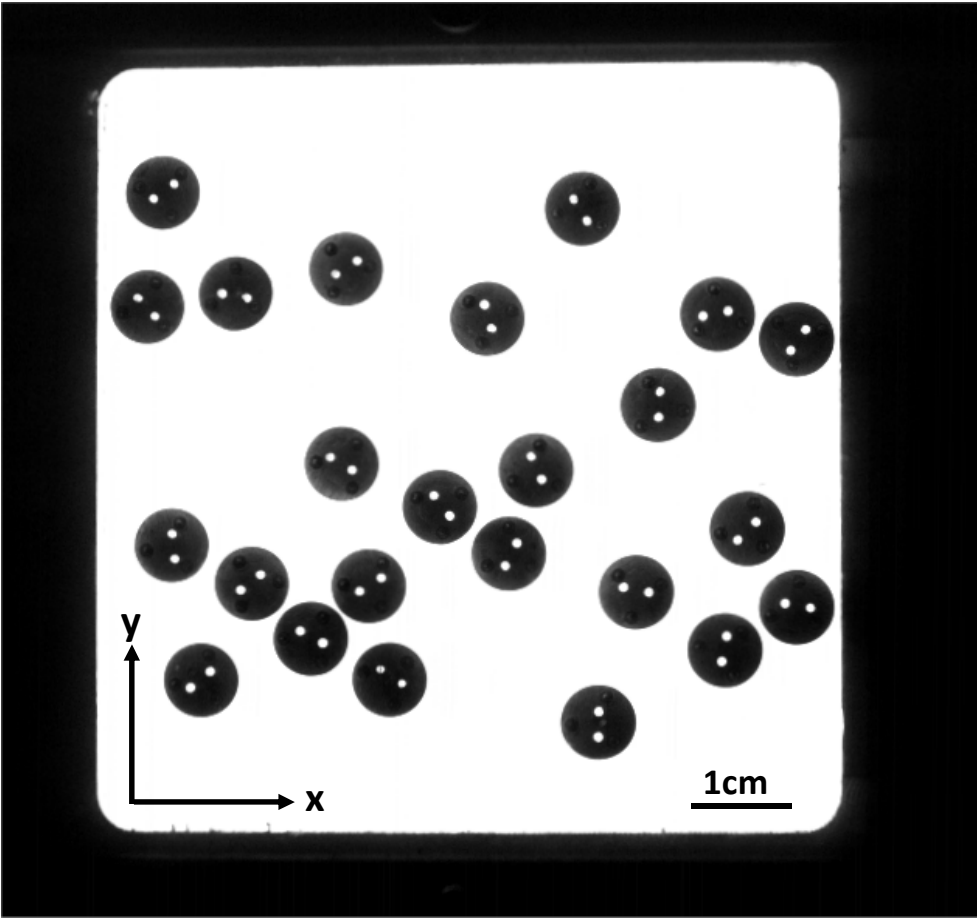

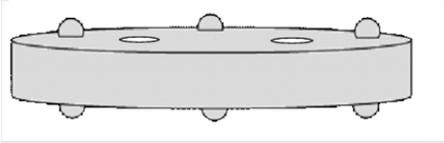 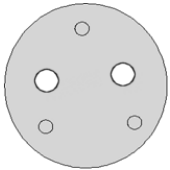

**Figure 3**

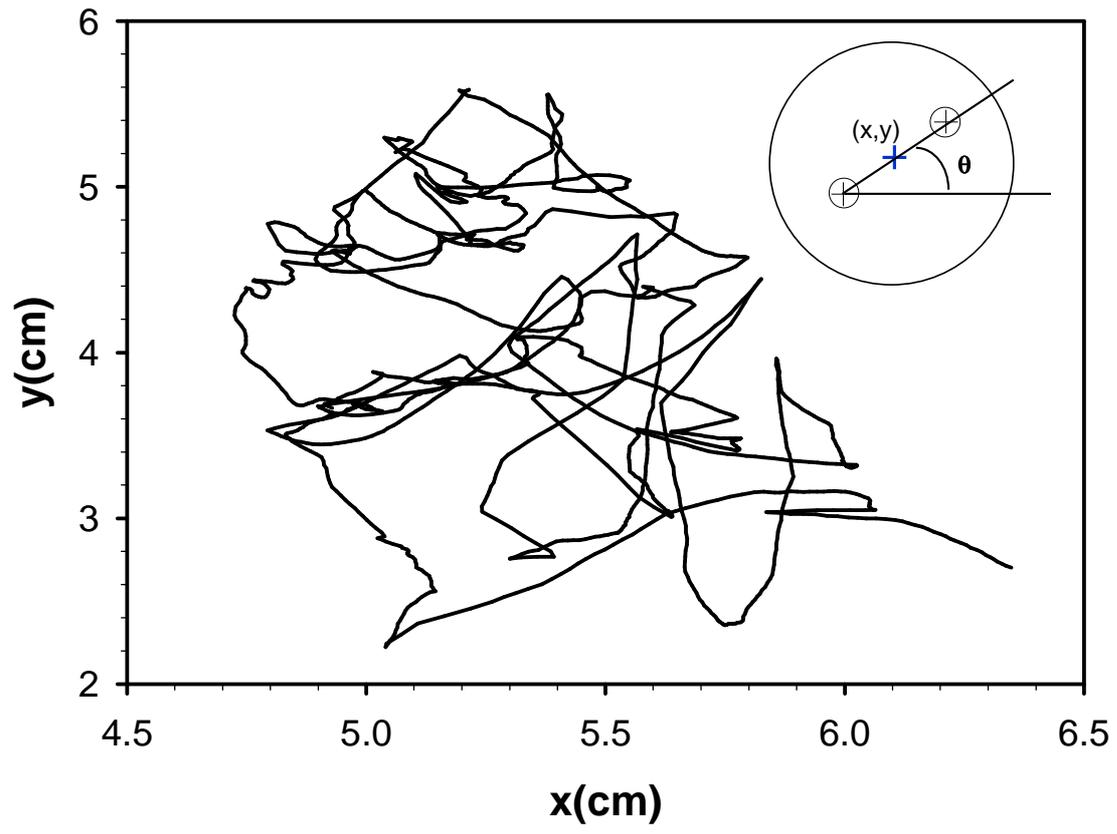

**Figure 4**

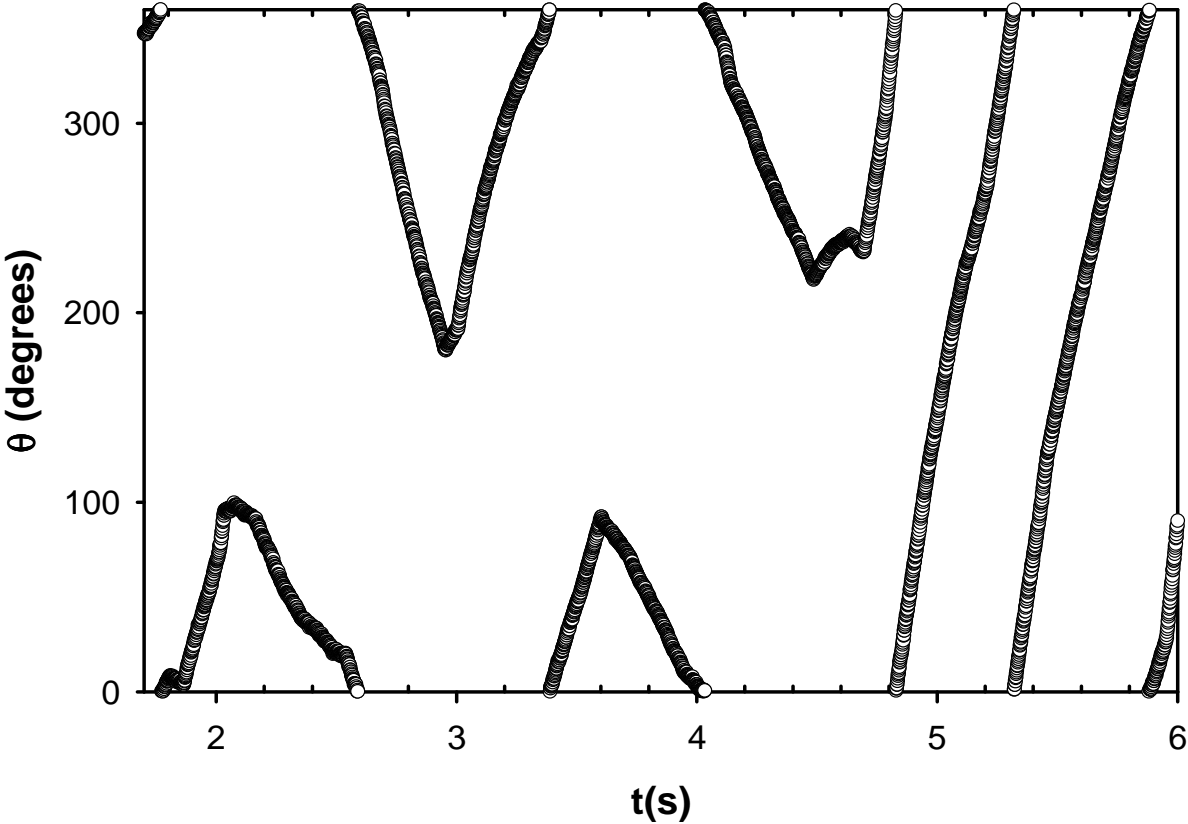

**Figure 5**

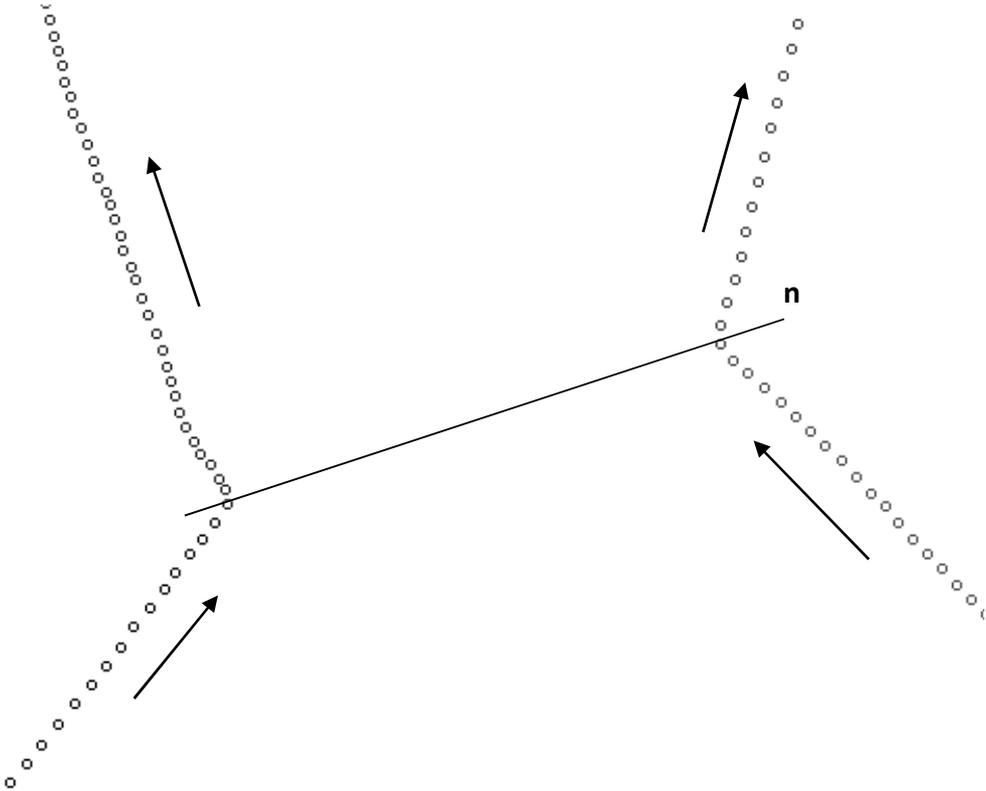

**Figure 6**

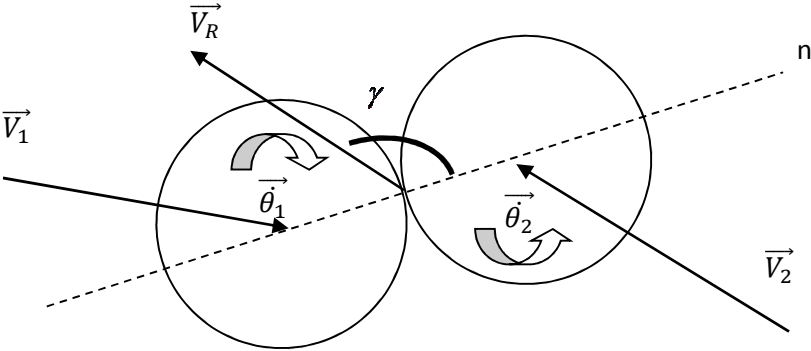

**Figure 7**

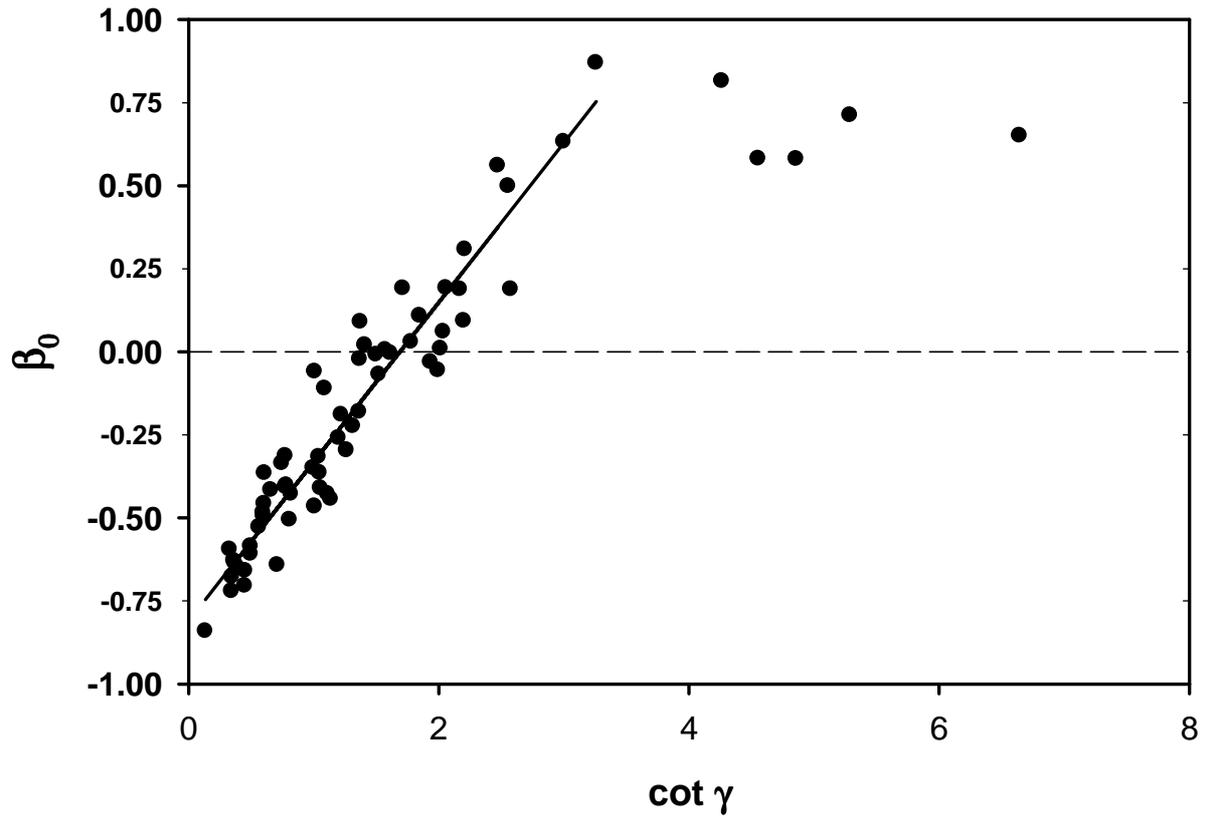

**Figure 8**

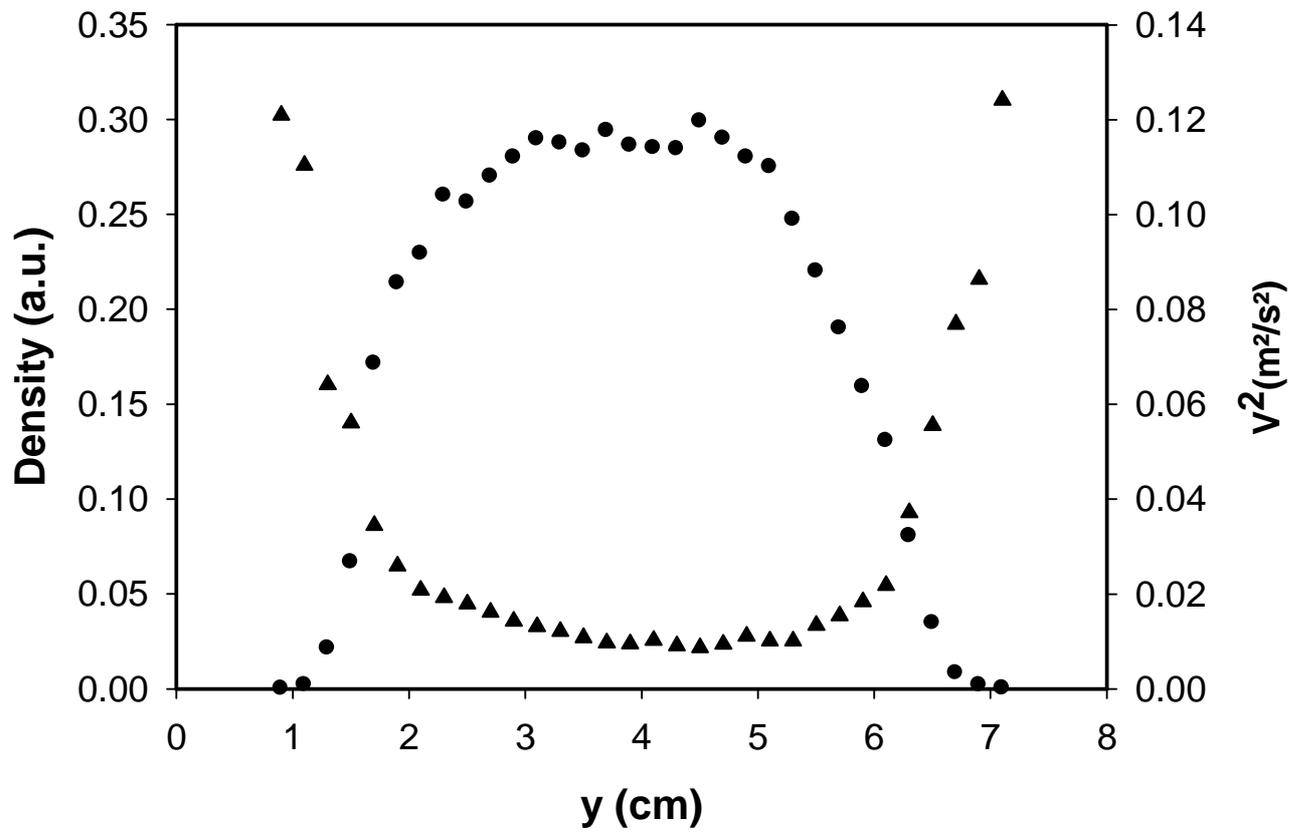

**Figure 9**

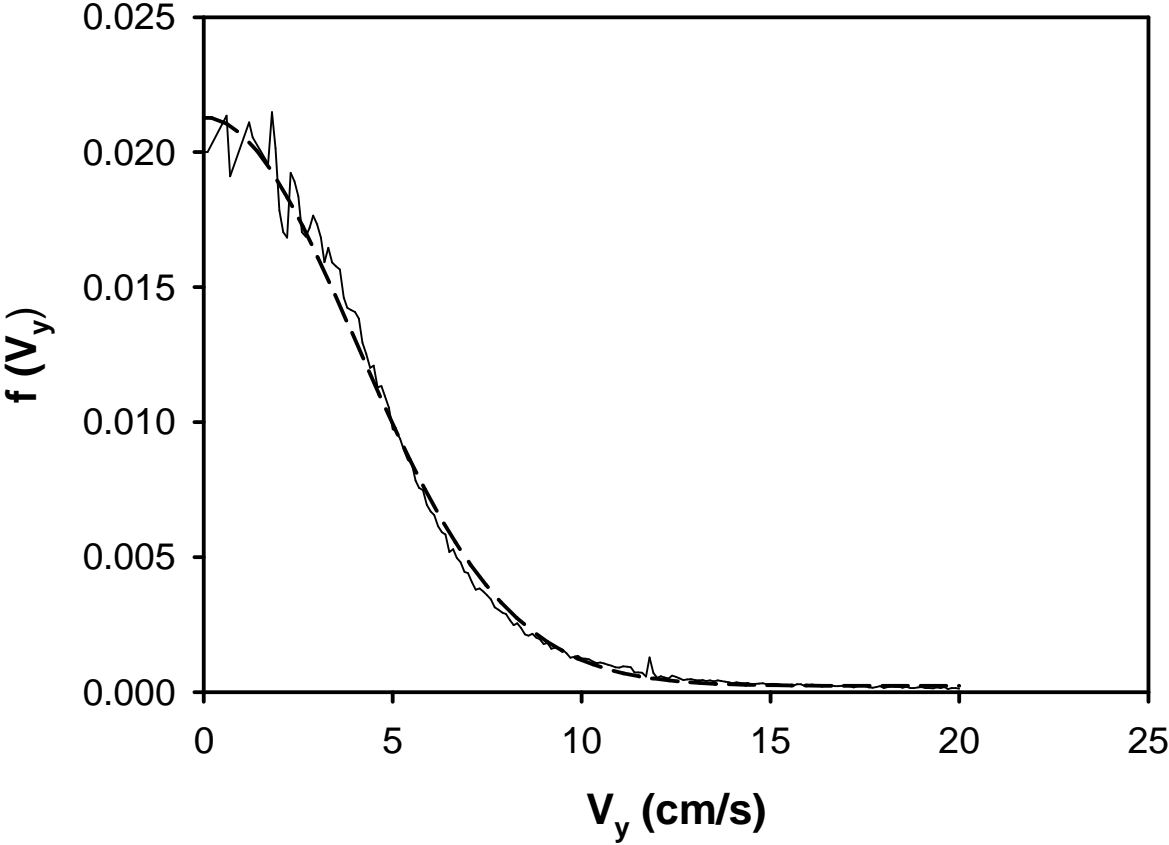

**Figure 10**

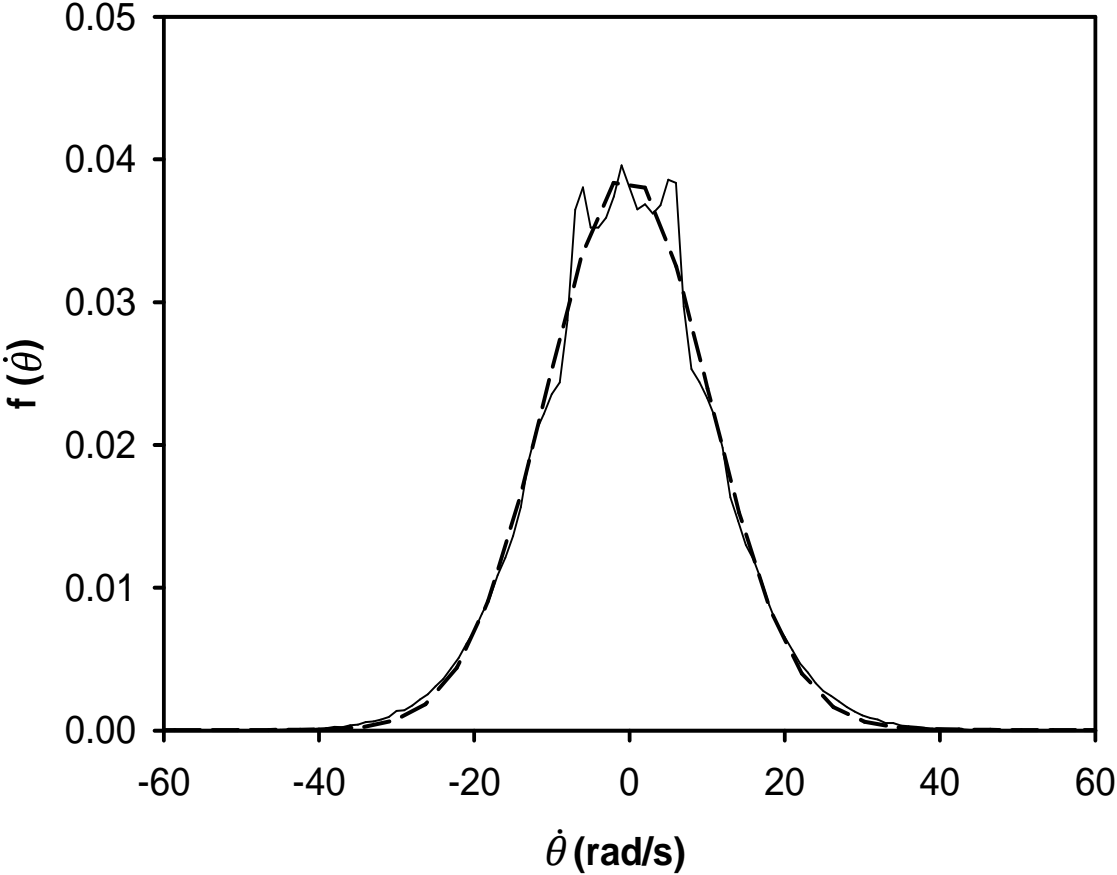

**Figure 11**

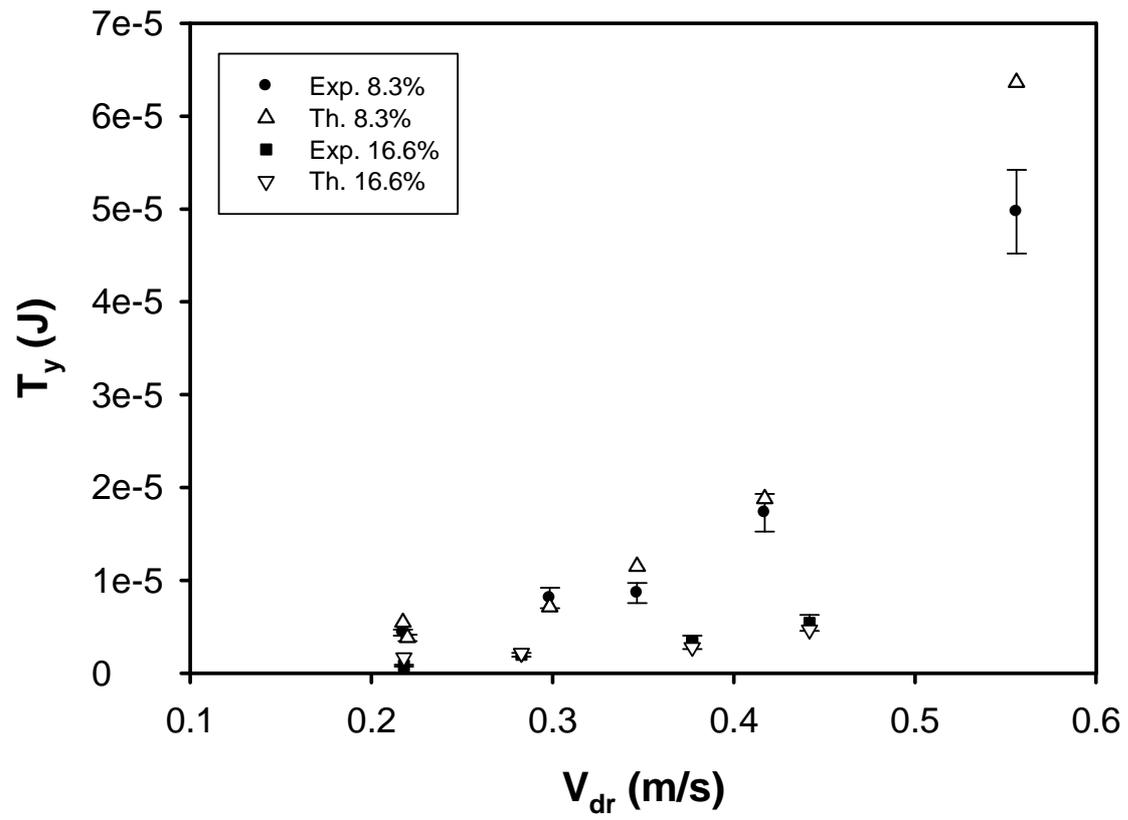